\documentclass[acmtog, screen=true]{acmart}  

\usepackage{preamble}

\acmSubmissionID{466}

\copyrightyear{2024}
\acmYear{2024}
\setcopyright{acmlicensed}
\acmConference[SA Conference Papers '24]{SIGGRAPH Asia 2024 Conference Papers}{December 3--6, 2024}{Tokyo, Japan}
\acmBooktitle{SIGGRAPH Asia 2024 Conference Papers (SA Conference Papers '24), December 3--6, 2024, Tokyo, Japan}
\acmDOI{10.1145/3680528.3687608}
\acmISBN{979-8-4007-1131-2/24/12}

\begin{document}

\title{\name: Differentiable CSG via Rasterization}

\author{Haocheng Yuan}
\orcid{0009-0007-1717-1585}
\affiliation{%
 \institution{The University of Edinburgh}
 \streetaddress{10 Crichton Street}
 \city{Edinburgh}
 \country{United Kingdom}
 }
\email{H.C.Yuan@ed.ac.uk}

\author{Adrien Bousseau}
\orcid{0000-0002-8003-9575}
\affiliation{%
 \institution{Inria, Universit\'{e} C\^{o}te d'Azur}
 \streetaddress{2004 route des lucioles}
 \city{Valbonne}
 \country{France}
}
\email{adrien.bousseau@inria.fr}

\author{Hao Pan}
\orcid{0000-0003-3628-9777}
\affiliation{%
 \institution{Microsoft Research Asia}
 \streetaddress{No.5 Danling Rd}
 \city{Beijing}
 \country{China}
}
\email{haopan@microsoft.com}

\author{Chengquan Zhang}
\orcid{0009-0000-7083-3671}
\affiliation{%
 \institution{Nanjing University}
 \streetaddress{163 Xianlin Rd}
 \city{Nanjing}
 \country{China}
 }
\email{zhangandresnju@gmail.com}

\author{Niloy J. Mitra}
\orcid{0000-0002-2597-0914}
\affiliation{%
\institution{University College London, Adobe Research}
\streetaddress{169 Euston Square}
\city{London}
\country{United Kingdom}
}
\email{n.mitra@cs.ucl.ac.uk}

\author{Changjian Li}
\orcid{0000-0003-0448-4957}
\affiliation{%
 \institution{The University of Edinburgh}
 \streetaddress{10 Crichton Street}
 \city{Edinburgh}
 \country{United Kingdom}
 }
 \email{Changjian.li@ed.ac.uk}


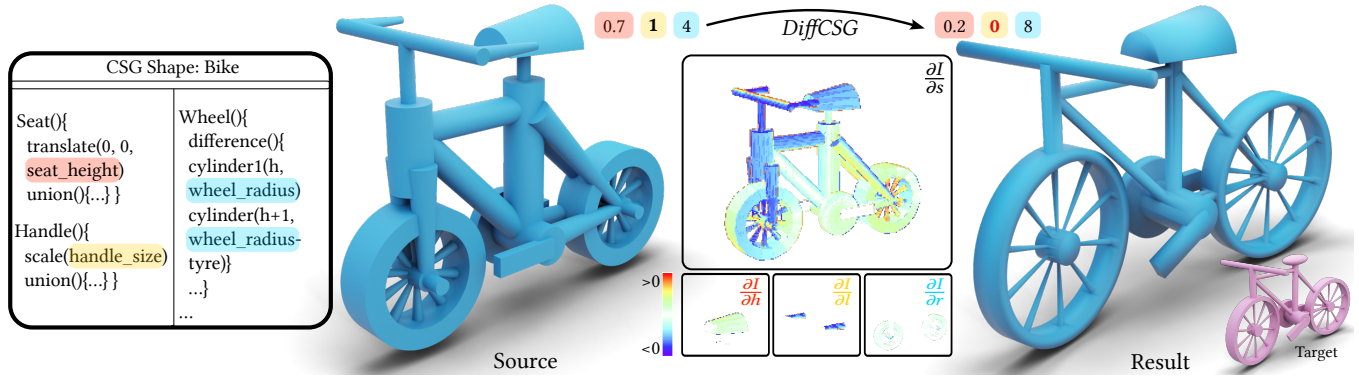
\begin{teaserfigure}
\centering
\begin{overpic}[width=\textwidth]{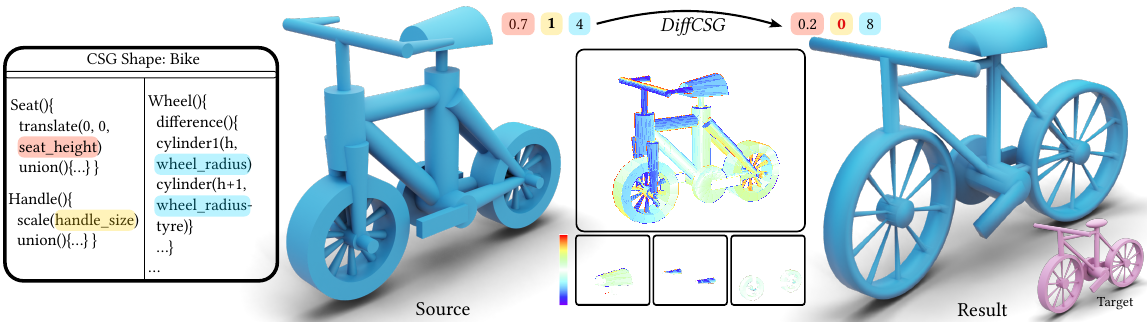}
    \put(68,21.5) {$\pdv{I}{s}$}
    \put(54.3, 5.7) {\small \textcolor{myorange}{$\pdv{I}{h}$}}
    \put(61.2, 5.7) {\small \textcolor{myyellow}{$\pdv{I}{l}$}}
    \put(68, 5.7) {\small \textcolor{myblue}{$\pdv{I}{r}$}}
    \put(47.2, 6.5) {\footnotesize >$0$}
    \put(47.2, 1.5) {\footnotesize <$0$}
\end{overpic}
\caption{\textbf{Differentiable CSG optimization.} 
Given the CSG model of a bike (left), our \name renders the corresponding shape in a differentiable manner, such that its continuous parameters can be optimized to best match multi-view renderings of a target shape (right, pink). Our solution builds upon differentiable rasterization to compute image gradients with respect to CSG parameters (middle). Here we visualize the per-pixel gradient contribution for the global scale parameter $s$, the \textcolor{myorange}{seat height $h$}, the \textcolor{myyellow}{handle size $l$}, and the \textcolor{myblue}{wheel radius $r$} (we cropped the gradient visualizations around the areas of interest for $h$, $l$ and $r$). In this example, the optimization decreased the seat height, increased the wheel radius, and made the handles vanish by setting their size to $0$ (top middle). The optimization also adjusted the orientation of the pedals.
}
\label{fig:teaser}
\end{teaserfigure}

\begin{abstract}
Differentiable rendering is a key ingredient for inverse rendering and machine learning, as it allows to optimize scene parameters (shape, materials, lighting) to best fit target images. Differentiable rendering requires that each scene parameter relates to pixel values through differentiable operations. While 3D mesh rendering algorithms have been implemented in a differentiable way, these algorithms do not directly extend to Constructive-Solid-Geometry~(CSG), a popular parametric representation of shapes, because the underlying boolean operations are typically performed with complex black-box mesh-processing libraries.
We present an algorithm, \name, to render CSG models in a differentiable manner. Our algorithm builds upon CSG rasterization, which displays the result of boolean operations between primitives \textit{without} explicitly computing the resulting mesh and, as such,
bypasses black-box mesh processing.
We describe how to implement CSG rasterization within a differentiable rendering pipeline, taking special care to apply antialiasing along primitive intersections to obtain gradients in such critical areas. Our algorithm is simple and fast, can be easily incorporated into modern machine learning setups, and enables a range of applications for computer-aided design, including direct and image-based editing of CSG primitives.
\rev{Code and data: \href{https://yyyyyhc.github.io/DiffCSG/}{https://yyyyyhc.github.io/DiffCSG/}.}
\end{abstract}

%
%


\maketitle

\section{Introduction}
\label{sec:intro}

\begin{figure*}[!t]
    \centering
    \begin{overpic}[width=\linewidth]{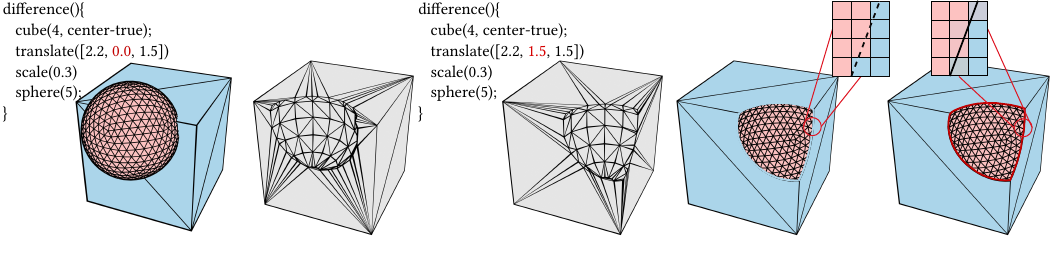}
        \put(8,0.5) {\small (a) Input CSG program}
        \put(27,0.5) {\small (b) Meshed CSG}
        \put(49.5,0.5) {\small (c) Meshed CSG}
        \put(45.5,-1.5) {\small for different parameter values}
        \put(66,0.5) {\small (d) Rasterized CSG}
        \put(85.5,0.5) {\small (e) Rasterized CSG}
        \put(81,-1.5) {\small with intersection edges antialiased}
    \end{overpic}
    \caption{\textbf{Ideation.} Given two intersecting primitives (a), the result of subtracting one primitive from the other can be obtained by computing boolean operations on meshes explicitly (b). However, the topology of the resulting mesh (number of vertices, connectivity) needs to be updated for any parameter change (b-c), which is costly and complex to implement in a differentiable way. The Goldfeather algorithm displays the result of the subtraction while maintaining the original primitive meshes (d), but the discontinuity formed along the intersection of the two primitives is not anti-aliased (d, inset). We detect intersection edges explicitly and apply anti-aliasing on them (e), which allows back-propagation of gradients from pixels to primitive triangles, all the way to primitive parameters. 
    }
    \label{fig:illustration}
\end{figure*}
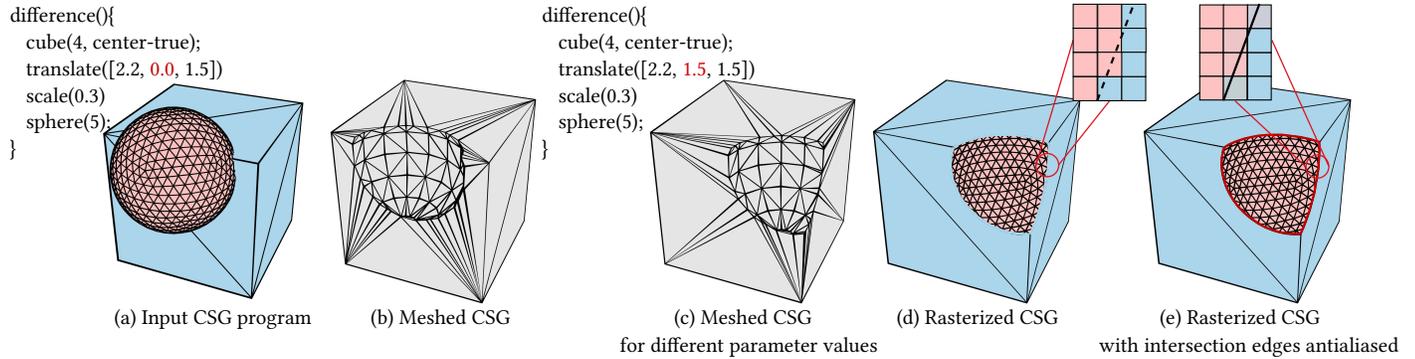

Constructive-Solid-Geometry (CSG) represents 3D shapes as parametric primitives (\eg cylinders, cuboids) combined with boolean operators (\eg union, intersection, difference) \cite{Voelcker1977}. CSG models are particularly popular in rapid prototyping as they allow designers to create editable shapes whose dimensions can later be adjusted to best fit different designs \cite{OpenSCAD,Customizer,Oehlberg2015}.

However, authoring and reusing CSG models requires programming skills, not only to define the primitives and boolean operators that form the shape's structure but also to adjust the individual parameters of all its components. 
Although specialized solutions exist to optimize certain CAD programs, they can only handle a small number of parameters \cite{michel2021dag}, or support unions of primitives but no intersections or differences \cite{cascaval2022differentiable,gaillard2022automatic,Kodnongbua2023}.
In contrast, differentiable rendering has recently gained popularity as a flexible way to optimize shape parameters to best fit target images \cite{Zhao2020}, enabling diverse applications like image-based shape reconstruction \cite{loubet2019reparameterizing} and text-based shape generation \cite{chen2023fantasia3d, liao2023tada}.
Unfortunately, existing differentiable renderers are not well suited to optimize parameters of CSG models. 

On the one hand, fast differentiable rasterizers work on meshes \cite{laine2020modular}. But computing boolean operations on meshed primitives is a complex and costly operation \cite{CPAL22}, that needs to be executed for any parameter update since such operations produce drastically different meshes for different parameter values (Fig. \ref{fig:illustration}b,c). Furthermore, boolean operations are typically implemented with mesh processing libraries that do not support automatic differentiation.
On the other hand, boolean operations can be expressed on Signed Distance Fields (SDFs) in a differentiable way \cite{kania2020ucsg}, but custom analytical SDF expressions need to be defined for each type of primitive.

We propose an efficient and flexible solution to optimize the \emph{continuous} parameters of CSG models using differentiable rendering.
On the one hand, we represent CSG models using boolean operators applied on meshes, which allows us to support any primitive that can be defined by a fixed tesselation, including the ones supported by popular CSG modelers \cite{OpenSCAD} (cuboid, cylinder, sphere, sketch-extrude). On the other hand, we do \textit{not} compute mesh boolean operations explicitly, avoiding the need to call complex black-box geometry processing libraries.
By maintaining the original tesselation of the primitives, we can backpropagate gradients from any point on the shape to the primitive vertices, all the way to the primitive parameters that control these vertices. 

We achieve this unique combination of features by revisiting CSG rasterization algorithms \cite{goldfeather1989near,stewart1998improved} under the new perspective of differentiable rendering. CSG rasterization performs depth tests against front and back faces of each primitive to deduce the fragments that form the rendered CSG shape (Fig.~\ref{fig:illustration}d, Fig.~\ref{fig:Goldfeather}). 
Then, differentiable rasterization provides a way to backpropagate gradients from fragments to mesh vertices \cite{laine2020modular}. However, care must be taken at discontinuities, such as across contours and silhouettes where the set of vertices affecting a pixel color change abruptly. In such cases, existing algorithms achieve differentiability by leveraging antialiasing to interpolate fragments across discontinuity edges \cite{li2018differentiable,laine2020modular}. 
But fast rasterizers only apply antialiasing along visibility edges detected on the input mesh --- this is insufficient for CSG rasterization as it also produces discontinuities at the \emph{intersection} of primitives. Since these discontinuities are not antialiased by state-of-the-art differentiable rasterization (Fig.~\ref{fig:illustration}d - inset), they do not provide valuable gradients. This lack of gradients is especially problematic for CSG optimization because primitive intersections are precisely among the visual features that vary most for different values of primitive parameters. 

Fortunately, while computing boolean operations on meshed primitives is a complex and costly procedure, detecting edges formed by intersections of primitives is a much easier and tractable task. Doing so in a differentiable manner allows us to feed additional edges to the anti-aliasing module and achieve differentiability over the entire CSG rendering pipeline (Fig.~\ref{fig:illustration}e - inset).
We demonstrate the simplicity and practicality of this approach by integrating it in Nvdiffrast \cite{laine2020modular}, a popular framework for differentiable rasterization. We then evaluate our CSG optimization on a benchmark of 50 CAD shapes and illustrate its potential on several downstream tasks, such as image-based editing and direct 3D shape manipulation.

\section{Related Work}
\label{sec:rw}

\begin{figure*}
    \centering
    \begin{overpic}[width=\linewidth]{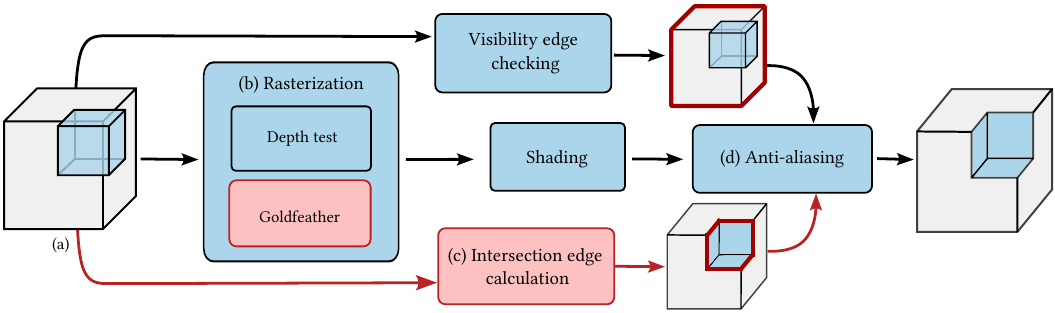}
    \end{overpic}
    \caption{\textbf{Algorithm Overview.} Our algorithm adapts a differentiable rasterization pipeline (light blue) to render CSG models (a) in a differentiable way. First, we replace the standard depth test by the Goldfeather algorithm (b), which selects among front and back faces of the CSG primitives the ones to be displayed according to boolean operations. Second, we detect intersection edges between CSG primitives (c) and provide these edges to the anti-aliasing module (d). Proper anti-aliasing is critical to allow back-propagation of gradients from the final image all the way to the primitive parameters. 
    }
    \label{fig:pipeline}
\end{figure*}

Modern Computer-Aided-Design (CAD) strongly relies on parametric representations of 3D shapes to encode variations of a design.
Constructive-Solid-Geometry (CSG) \cite{Voelcker1977} is one such representation where a design is defined both by discrete parameters (number and type of geometric primitives, type of boolean operators applied to these primitives) as well as by continuous parameters (position, orientation, and dimensions of the primitives). 
\rev{CSG is the core modeling metaphor of popular software like OpenSCAD \cite{OpenSCAD}, as well as a key component of feature-based CAD systems like OnShape \cite{onshape} and Fusion360 \cite{fusion360,Willis2021} that heavily rely on sketch-extrude operators to create or remove geometry.}
In this paper, we focus on application scenarios where the discrete structure of the CSG model is given, while its continuous parameters need to be adjusted. Such editing tasks frequently occur when users want to reuse an existing design \cite{Oehlberg2015}, but are challenging to perform as soon as the design exposes multiple parameters that are difficult to relate to the desired outcome \cite{gonzalez2024}. 
The ability to optimize continuous shape parameters can also benefit reverse-engineering tasks where the structure of the CSG model is discovered by other means\rev{, including methods based on program synthesis that typically quantize continuous parameters to a small set of integer values to treat all program tokens as discrete variables \cite{du2018inversecsg,Ellis2019, Sharma_2018_CVPR}. Our approach could be used to refine continuous parameters once the discrete structure has been discovered, or used in conjunction with iterative program synthesis \cite{kapur2024diffusion} to jointly solve for discrete and continuous parameters.}

Several approaches have been recently proposed for differentiable execution of CAD models, but most of them do not apply to CSG. \citet{cascaval2022differentiable} and \citet{gaillard2022automatic} use automatic differentiation to relate the position of mesh vertices to the shape parameters that control these vertices, which allows them to offer users a bidirectional workflow where CAD models can be edited either in parameter space or in mesh space. \citet{Kodnongbua2023} further combine differentiable CAD execution with differentiable rendering to fit CAD models to images. \rev{Similarly, \citet{Worchel2023} demonstrate how to extract triangle meshes from parametric surfaces (B\'{e}zier, B-spline, NURBS surfaces) in a differentiable way, enabling differentiable rendering of such surfaces}.
However, extending these systems to CSG would require implementing boolean operations on meshes within an automatic differentiation framework, a non-trivial task given the complexity of robust mesh boolean algorithms \cite{CPAL22}.
A notable exception is the work of \citet{Gonzalez2023}, which allows users to edit the position, rotation, and scale of CSG primitives directly in 3D space. But this system relies on instrumenting the CSG editor OpenSCAD \cite{OpenSCAD} with custom 3D widgets, rather than employing differentiable execution to automatically optimize arbitrary parameters. 

\citet{michel2021dag} propose a method to assign a unique identifier to any point of a parametric shape expressed by a direct acyclic graph (including CSG trees), enabling to compute pointwise derivatives by finite differences rather than through automatic differentiation. While this approach is well adapted to click-and-drag interactions, its computational cost limits it to models controlled by a small number of parameters. In contrast, our approach computes gradients with respect to tens of parameters, over all visible points of the object, enabling image-guided optimization of CSG parameters.

Given the difficulties induced by boolean operations on meshes, several authors perform machine learning with CSG models using alternative shape representations such as occupancy grids and signed distance functions (SDFs) \cite{kania2020ucsg,ren2021csg,Lambourne2022}, where boolean operations are expressed as min/max functions implemented with softmax to achieve differentiability. \rev{Recently, \citet{Liu2024} proposed to replace the min and max operators by a unified, fuzzy-logic operator that enables to also optimize the type of boolean operation applied on primitives.}
However, employing SDFs raises several challenges. First, the SDF of each primitive type needs to be defined analytically, as a function of the primitive parameters. While analytical SDFs have been defined for simple primitives such as spheres, cylinders and cuboids (see supplemental material from \cite{kania2020ucsg,ren2021csg}), expressing complex primitives, such as sweep surfaces formed by B\'{e}zier curves, would be more involved \cite{QuilezSDFs}. Second, evaluating SDFs throughout optimization is costly, as demonstrated in our evaluation (see Sec. \ref{subsec:comp}). Finally, optimizing SDFs according to image-based losses requires rendering them with differentiable sphere tracing, which is also a costly procedure \cite{Vicini2022sdf,Bangaru2022NeuralSDFReparam}.
Our solution avoids these challenges by working with meshes, which are easy to define for diverse parametric primitives and are fast to rasterize. Yet, 
we avoid computing boolean operations on meshes explicitly and instead leverage differentiable rasterization to backpropagate image-space gradients to primitive vertices, all the way to primitive parameters.

\section{Method}
\label{sec:preliminary}

Figure~\ref{fig:pipeline} provides an overview of our algorithm.
Our method takes as input a CSG model composed of a set of parametric primitives combined with boolean operations (Fig.~\ref{fig:pipeline}a). We build on the \rev{Goldfeather algorithm \shortcite{goldfeather1989near}}, which performs visibility tests during rasterization to render CSG models without computing boolean operations on meshes explicitly. While we can perform these visibility tests in place of the standard depth test in a differentiable rasterization pipeline (Fig.~\ref{fig:pipeline}b), the final image contains edges not present in the original geometry. We detect these edges explicitly by finding intersections between the parametric primitives (Fig.~\ref{fig:pipeline}c). We then provide the detected intersection edges along with visibility edges to the anti-aliasing module of a differentiable rasterization pipeline (Fig.~\ref{fig:pipeline}d), which allows back-propagation of image-space gradients to the respective primitives.

We now provide background details about the CSG rasterization and differentiable rasterization algorithms we build upon, before describing how we combine these two components with intersection edge detection to achieve differentiable CSG.

\subsection{Background on CSG Rasterization}
\label{sec:preliminaryCSG}
Fig. \ref{fig:pipeline} (light blue) depicts the main stages of a modern rasterization pipeline \cite{laine2020modular}. Starting from input triangle meshes, the rasterization module projects the triangles on the screen and uses the depth buffer to test which fragments are visible. These fragments are subsequently shaded and anti-aliased.

\begin{figure}[!tb]
    \centering
    \begin{overpic}[width=\linewidth]{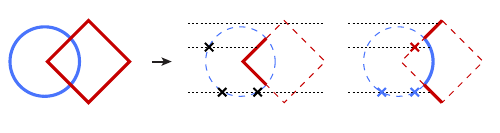}
        \put(1,0) {\small (a) CSG primitives}
        \put(44,0) {\small (b) $\textcolor{blue}{A} \cap \textcolor{red}{B}$}
        \put(78,0) {\small (c) $\textcolor{red}{B} - \textcolor{blue}{A}$}
        \put(8,10) {\small \textcolor{blue}{A}}
        \put(17,17) {\small \textcolor{red}{B}}
        \put(28.5,16) {\small view}
    \end{overpic}
    \caption{\textbf{Goldfeather algorithm.} The algorithm takes as input individual primitives of the CSG model (a). For an intersection, only the front-facing fragments that are occluded by an odd number of polygons are displayed (b, crosses depict occlusions along viewing rays). 
    For a subtraction, the two primitives apply different parity tests against each other (c). Considering $\textcolor{red}{B} - \textcolor{blue}{A}$, the algorithm displays the front-facing fragments of $\textcolor{red}{B}$ that are occluded by an even number of polygons from $\textcolor{blue}{A}$, and the back-facing fragments of $\textcolor{blue}{A}$ that are occluded by an odd number of polygons from $\textcolor{red}{B}$.
    }
    \label{fig:Goldfeather}
\end{figure}
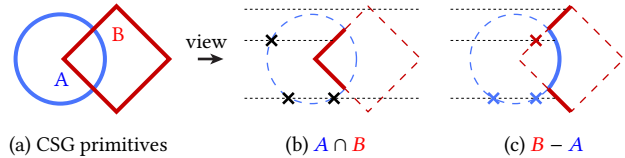

The Goldfeather algorithm \shortcite{goldfeather1989near} is a simple modification of this pipeline (Fig. \ref{fig:pipeline}(b)), which consists of replacing the depth test with a combination of depth and parity tests that implements the boolean operations applied on primitives. 
Fig.~\ref{fig:Goldfeather} illustrates the algorithm on two convex primitives, $A$ and $B$.
To display the intersection of the two primitives, the algorithm checks whether there is an odd number of polygons lying between the front faces of the primitives and the camera (Fig.~\ref{fig:Goldfeather}(b)). 
For the subtraction of A from B, the algorithm keeps the fragments from back faces of A that have an odd number of polygons from B occluding them from the camera, and keeps the fragments from front faces of B that have an even number of occluding polygons from A (Fig.~\ref{fig:Goldfeather}(c)). 
For a union of primitives, the standard depth test is applied. 
We refer the reader to \cite{goldfeather1989near,stewart1998improved,kirsch2005opencsg} for additional details, such as the handling of non-convex primitives and algorithm optimizations. 

\subsection{Background on Differentiable Rasterization}
Differentiable rendering aims at computing how much the pixel values of an image change in reaction to small changes in scene parameters (geometry, material, light, camera). Differentiable rendering is a critical ingredient of inverse rendering, which aims at optimizing scene parameters to best match a target image. In the past few years, significant progress has been made in differentiable rendering based on either \rev{raytracing \cite{Zhao2020,Zhang2020PSDR,bangaru2020warpedsampling,Zhang2023Projective} or rasterization \cite{liu2019soft,laine2020modular,cole2021differentiable}}. We focus on rasterization for its real-time \rev{performance}, and extend it to CSG models for differentiable CSG rendering.

We build our algorithm on the differentiable rasterizer of \citet{laine2020modular}, i.e. Nvdiffrast, which implements the main steps of the rasterization pipeline in a differentiable way. Several steps of this pipeline are continuous by construction, such as barycentric interpolation of vertex attributes, or trilinear interpolation of texture values, and as such easy to differentiate. In contrast, the visibility test creates sharp discontinuities in the image, which provide no gradients for vertex positions. Similar to \citet{li2018differentiable}, Laine et al. apply screen-space anti-aliasing along visibility edges to obtain smooth discontinuities (Fig. \ref{fig:pipeline}(d)), where pixels crossed by silhouette edges are blended with their neighbors, with weights determined by edge coverage (see Fig. \ref{fig:pixel_antialiasing}). 
However, Laine et al. only consider edges from the input triangle mesh, and only apply anti-aliasing on the ones that correspond to silhouettes. This procedure is not sufficient to obtain proper gradients for the Goldfeather algorithm, since this algorithm introduces sharp discontinuities along primitive intersections, which can occur away from silhouettes and do not necessarily align with the input mesh edges.

\subsection{Differentiable Rasterization of CSG Models}
\label{subsec:dr_csg}

\paragraph{Differentiation through primitive intersections}
Our key idea is to make the Goldfeather algorithm compatible with differentiable rasterization by explicitly detecting and antialiasing intersection edges between CSG primitives.
To do so, we consider all pairs of triangles from different primitives and express their intersections as a function of the triangle vertices (see \emph{supplementary} for technical details). We perform this computation in parallel using PyTorch for both differentiability and minimal overhead. 
To tell which pixels are crossed by visible intersection edges, we then render the detected intersection edges into pixels using OpenGL with depth test. 
Finally, we send these pixels along with the endpoints of their crossing edges to the anti-aliasing module for pixel blending, as is done for silhouette edges. 

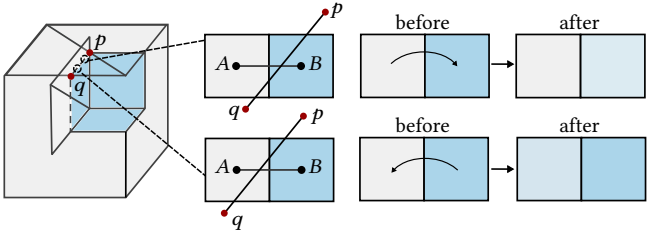
\begin{figure}[!b]
    \centering
    \begin{overpic}[width=\linewidth]{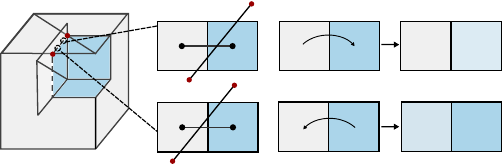} 
        \put(14.2, 27) {\small $p$}
        \put(11, 19.8) {\small $q$}
        \put(61, 29.5) {\small before}
        \put(86.5, 29.5) {\small after}
        \put(61, 13.5) {\small before}
        \put(86.5, 13.5) {\small after}
        \put(51, 32) {\small $p$}
        \put(35, 16) {\small $q$}
        \put(48, 15.5) {\small $p$}
        \put(35.5, -1) {\small $q$}
        \put(33, 23) {\small $A$}
        \put(33, 7) {\small $A$}
        \put(47.5, 23) {\small $B$}
        \put(47.5, 7) {\small $B$}
    \end{overpic}
    \caption{\textbf{Pixel anti-aliasing.} 
    \rev{On the left, an intersection edge $(p,q)$ is formed by the top face of the gray cube and the vertical face of the blue cube. The two endpoints of the intersection edge are colored red.
    On the right, when considering the color blending of the crossing pixels (i.e., the gray pixel $A$ and the blue pixel $B$) of the intersection edge, two typical options may apply depending on which pixel is most covered by the edge.}
    In the top case, edge \rev{$(p,q)$} intersects the segment connecting centers of $A, B$ inside pixel $B$, which leads to the color of $A$ blending into $B$, \ie Color$^\textrm{B}_{\textrm{after}}$ = $\alpha$ * Color$^\textrm{A}$ + (1-$\alpha$)* Color$^\textrm{B}_{\textrm{before}}$. In the bottom case, the edge covers $A$ the most, so the color of $B$ is blended into $A$, \ie Color$^\textrm{A}_{\textrm{after}}$ = $\alpha$ * Color$^\textrm{B}$ + (1-$\alpha$) * Color$^\textrm{A}_{\textrm{before}}$. The blending weight $\alpha$ is a linear function of the location of the crossing point, from zero at the midpoint to $0.5$ at the pixel center.
    }
    \label{fig:pixel_antialiasing}
\end{figure}

Fig. \ref{fig:pixel_antialiasing} depicts the anti-aliasing calculation, which we adapted from \citet{laine2020modular} to account for intersection edges. In this calculation, pixel blending is a function of the endpoints of intersection edges, which we themselves express as a function of primitive vertices, allowing gradient back-propagation from pixels to intersection points and from intersection points to primitive vertices via the automatic differentiation mechanism in Nvdiffrast and PyTorch.

\paragraph{Back-propagation to primitive parameters}
The above differentiable rasterization allows us to back-propagate gradients from image pixels to mesh vertices. We next need to back-propagate gradients from vertices to primitive parameters. 
While one could express each vertex of the primitive mesh as a function of primitive parameters, this solution makes the implementation of each primitive tessellation quite cumbersome. We ease implementation of some of our primitives by binding vertices to the corresponding parameters via intrinsic scaling factors. For example, we express the radii (resp. height) of a tapered cylinder as horizontal (resp. vertical) scaling of its vertex positions, respectively. 
An exception is the B\'{e}zier curve, for which the polynomial expression that dictates vertex positions is a function of the curve control points.
We then express the extrinsic primitive parameters (position, orientation, uniform scale) as additional transformation matrices.

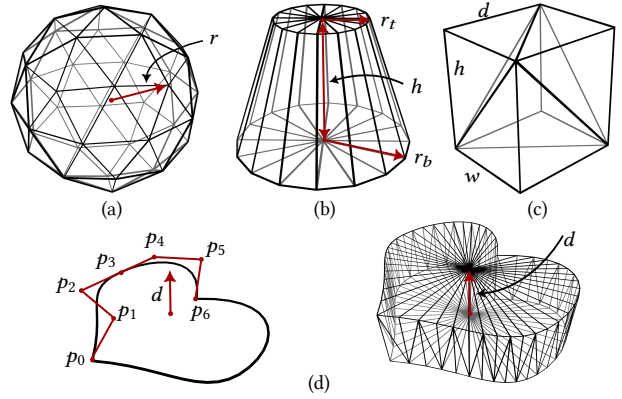
\begin{figure}[!tb]
    \centering
    \begin{overpic}[width=0.95\linewidth]{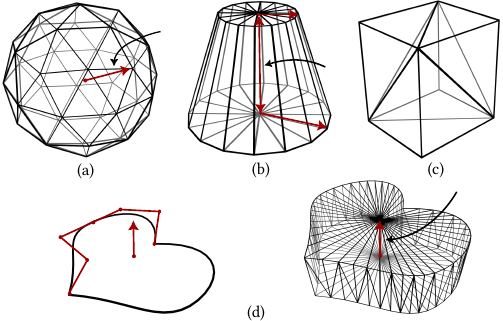} 
        \put(33, 57) {\small $r$}
        \put(61, 60) {\small $r_{t}$}
        \put(66.5, 37.5) {\small $r_{b}$}
        \put(66, 49) {\small $h$}
        \put(74.5, 34.5) {\small $w$}
        \put(73, 52) {\small $h$}
        \put(77, 61.5) {\small $d$}
        \put(23.5, 15) {\small $d$}
        \put(9.5, 5) {\small $p_0$}
        \put(18, 12.5) {\small $p_1$}
        \put(8, 17) {\small $p_2$}
        \put(14.5, 21) {\small $p_3$}
        \put(22.5, 24.5) {\small $p_4$}
        \put(32.5, 23) {\small $p_5$}
        \put(30, 12.5) {\small $p_6$}
        \put(91, 24) {\small $d$}
    \end{overpic}
    \caption{{\bf Primitive tessellation}. For sphere (a), the only parameter is the radius $r$. For cylinder (b), we sample $16$ points along the top and bottom circles to form the triangulation. Other than the height parameter $h$, we support \emph{taper}, \ie the top radius $r_t$ and the bottom radius $r_b$ can be different. Box (c) is the most simple primitive. We employ 12 triangles in the mesh and expose the height $h$, width $w$, and depth $d$ as the primitive parameters. For sketch-extrude (d), we resort to piecewise Bezier curves to represent the sketch (d, left) and sample $15$ points on each Bezier curve to form the triangulation after extrusion (d, right). In this case, all the control point positions (half of them are drawn) and the extrude distance $h$ are parameters.}
    \label{fig:primitive_tesse}
\end{figure}

\paragraph{Primitive tessellation}
Our method can handle arbitrary primitives with fixed tessellations. 
For proof of concept, we implemented the most common and expressive primitives shown in Fig. \ref{fig:primitive_tesse}.
We adopted a coarse tessellation of the primitives to favor efficiency over shape smoothness.

\subsection{Gradient-based Optimization of CSG Parameters}
\label{subsec:optimization}
Our algorithm is highly modular and can be plugged into different inverse-rendering applications based on gradient descent. In a typical optimization protocol, given a pre-defined CSG model with primitive parameters $\{\theta_i\}$, a target image $I_{\text{target}}$, and camera parameters $p_{\text{cam}}$, we optimize the parameters to achieve the target:
\begin{equation}
    \operatorname*{arg\,min}_{\{\theta_i\}} \mathcal{L} = ||I\left(\{\theta_i\}, p_{\text{cam}}\right) -  I_{\text{target}}||,
\end{equation}
where $I(\{\theta_i\}, p_{\text{cam}})$ is the rendered image from our \name, and the loss can be pixel-wise $L_1$ or $L_2$ loss. When multi-view target images with corresponding camera parameters are given, we extend the optimization by averaging over all views. 
Depending on applications, we experimented with target images representing the per-pixel normal of the target shape, or the solid color of its constituent primitives (Sec.~\ref{sec:results}). We used an image resolution of $512$x$512$ for all our experiments.

We implemented our optimization using PyTorch and the Adam optimizer \cite{kingma2014adam} with a default learning rate $1e^{-3}$. We set the loss threshold to $5e^{-4}$ regardless of whether color or normal images are used for the target, and the maximum number of optimization steps to $5000$. We terminate the optimization as soon as one of these conditions is reached. These three hyper-parameters can vary when applying the optimization to different applications.

\section{Results and Discussion}
\label{sec:results}

\subsection{Benchmark Evaluation}
\label{subsec:dataset_evaluation}

We provide as supplemental materials an evaluation of our algorithm on a benchmark of 50 CSG models. For each model, we applied random perturbations of primitive parameters to obtain source shapes that we then optimize to best align with the shape produced by the default parameters. This evaluation reveals that the success rate of our optimization depends on the magnitude of perturbation, and that models that are equipped with a few hyper-parameters are easier to optimize than the ones exposing many independent parameters. Please refer to supplemental materials for details about the composition of the benchmark and optimization results.

\begin{figure}[!htb]
    \centering    
    \includegraphics[width=0.99\linewidth]{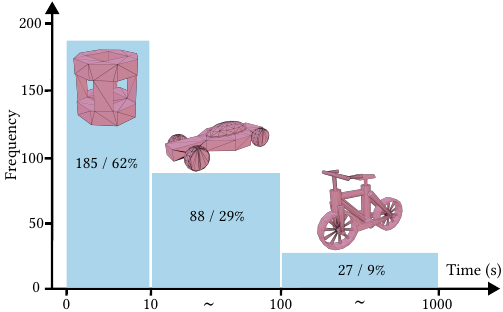}
    \caption{\textbf{Running time overview and typical cases.} Among all $300$ testing cases, the majority of the shapes ($62\%$) can be optimized within $10s$, while complex shapes like the bike can take up to $1000s$ for the optimization. Typical examples for each category are shown.}
    \label{fig:timeOverview}
\end{figure}

In terms of performance, $62\%$ of shapes required less than $10s$ to fit the target, $29\%$ required less than $100s$, and only a small percentage ($9\%$) of complex shapes required more than 1 minute but less than 20. Typical examples of these three categories are shown in Fig. \ref{fig:timeOverview}.
Table \ref{tab:runningTime_details} details the running time for representative shapes shown in the paper. 
The Goldfeather algorithm and edge calculation times mostly depend on the number of triangles composing the shape, while the optimization time is affected by both the number of triangles and the number of optimizable parameters. 
Note that our implementation of both the Goldfeather algorithm and edge calculation is not optimized; these steps could be more efficient. \rev{Please refer to the supplementary for a detailed discussion of scalability.}

\begin{table}[!htb]
    \centering
    \caption{\textbf{Running time breakdown for one iteration.} For each shape, we indicate the number of triangles and number of optimizable parameters, along with the time spent on Goldfeather algorithm, intersection edge detection, and optimization (in seconds). 
    }
    \label{tab:runningTime_details}
    \renewcommand{\arraystretch}{1.2}
    \resizebox{\linewidth}{!}{
    \begin{tabular}{r c c c c c c}
    \toprule[1pt]
    & \#Tri. & \#Para. & Goldf. & EdgeC. & Opt. & Total \\
    
    \cmidrule(lr){1-1}
    \cmidrule(lr){2-3}
    \cmidrule(lr){4-4}
    \cmidrule(lr){5-5}
    \cmidrule(lr){6-6}
    \cmidrule(lr){7-7}
    
    Fig.~\ref{fig:teaser}, Bike & 2980 & 17& 0.167 &0.072 &0.106 &0.345 \\ 
    \rowcolor{Gray}
    Fig.~\ref{fig:gallery}, RaceCar & 4336 & 22 & 0.163 &0.150&0.244 &0.557 \\ 
    Fig.~\ref{fig:gallery}, Moon & 128 & 3 & 0.019 &0.014 &0.028 &0.061 \\ 
    \rowcolor{Gray}
    Fig.~\ref{fig:gallery}, Nut & 88 & 16 &0.022 &0.012 &0.026 &0.060 \\ 

    \bottomrule[1pt]
    \end{tabular}
}
\end{table}

\subsection{Ablation Study}
\label{subsec:abl_study}

We first demonstrate the importance of anti-aliasing intersection edges to obtain accurate gradients. We then evaluate the impact of different types of interior signals that we use as targets for the optimization.

\begin{figure*}[t]
    \centering
    \begin{overpic}[width=0.99\textwidth]{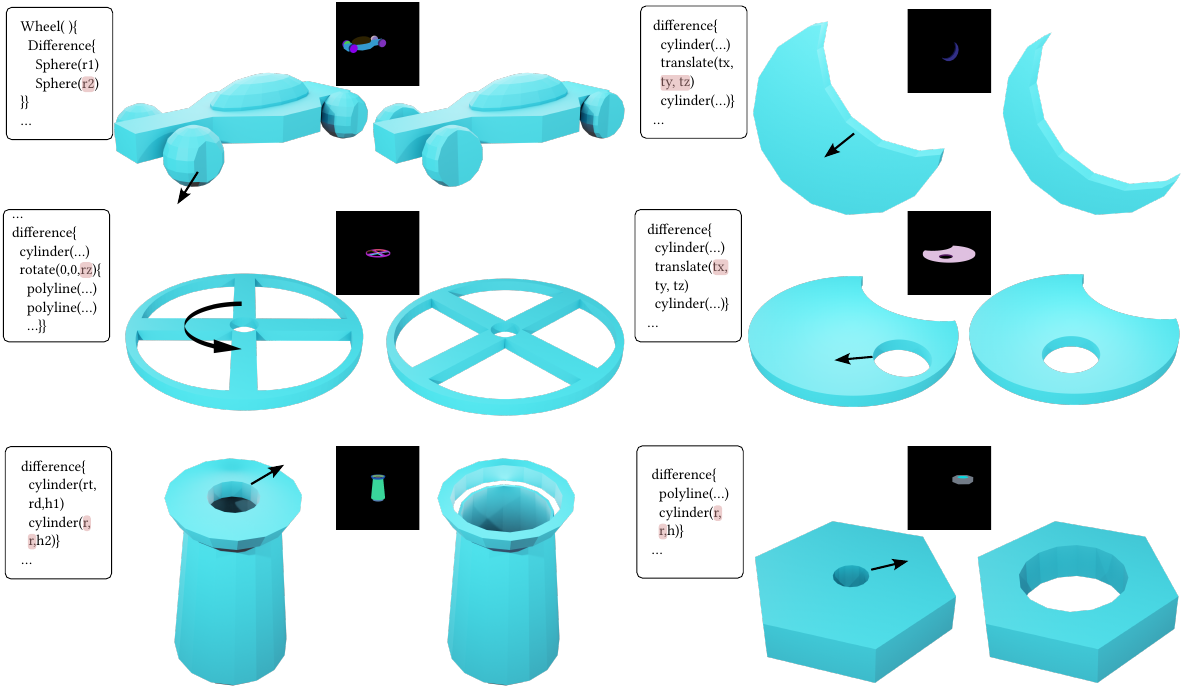}
    \end{overpic}
    \caption{\textbf{Visual Results.} Examples of shapes from our benchmark that require proper treatment of intersection edges to be optimized. Without our anti-aliasing, the optimization remains stuck in its initial state due to the discontinuities introduced by the difference operator. In each example, the source shape, resultant shape, and one of the target images rendered with per-primitive colors are shown. Black arrows indicate the desired change between the source and the target.
    \label{fig:gallery}
    }
\end{figure*}

\begin{figure}[!tb]
    \centering
    \begin{overpic}[width=0.99\linewidth]{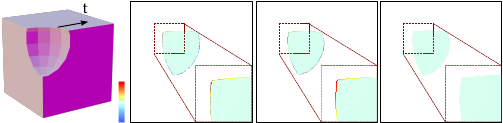}
         \put(30,-3.5) {\footnotesize \rev{(a) Finite diff.}}
         \put(52,-3.5) {\footnotesize (b) w/ anti-aliasing}
         \put(76.5,-3.5) {\footnotesize (c) w/o anti-alising}
         \put(45,20) {$\pdv{I}{t}$}
         \put(70,20) {$\pdv{I}{t}$}
         \put(95,20) {$\pdv{I}{t}$}
         \put(23.3,-2) {\scriptsize <$0$}
         \put(23.3,8.7) {\scriptsize >$0$}
    \end{overpic}
     \caption{\textbf{Gradient visualization.} We used finite differences to compute ground-truth image gradients for the translation $t$ of the subtracted sphere (a). Our anti-aliasing produces similar gradients along the intersection edges (b). Without anti-aliasing, the gradient information is incorrect (c). 
     }
    \label{fig:gradients}
\end{figure}

\begin{figure}[!tb]
    \centering
    \includegraphics[width=0.98\linewidth]{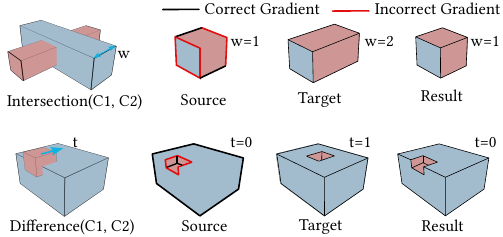}
    \caption{\textbf{Ablation study: impact of intersection edges.} Without our detection of intersection edges, the gradient information along intersections does not offer valuable information for optimization. In these two examples, the optimization remains stuck at its initial state because only the intersection edges indicate that the box should enlarge (top) or translate (bottom). Our full method, which includes intersection edges, succeeds in both cases.
    }
    \label{fig:abl_importance}
\end{figure}

\paragraph{Importance of intersection edges}
Fig. \ref{fig:abl_importance} illustrates the importance of intersection edges for CSG optimization in two didactic examples where only the intersection edges provide information about the primitive parameter to be optimized. If no anti-aliasing is applied along these edges, the optimization does not progress.
Fig. \ref{fig:gallery} illustrates more complex and realistic shapes that require correct gradients along intersections to be properly optimized. 
Furthermore, Fig. \ref{fig:gradients} provides a visual comparison between our gradient map, the gradient map obtained without anti-aliasing intersection edges, and a reference gradient map obtained with finite differences. This comparison highlights the inaccuracy of gradients along intersections if anti-aliasing is not properly applied.

\begin{figure}[!tb]
    \centering
    \includegraphics[width=0.96\linewidth]{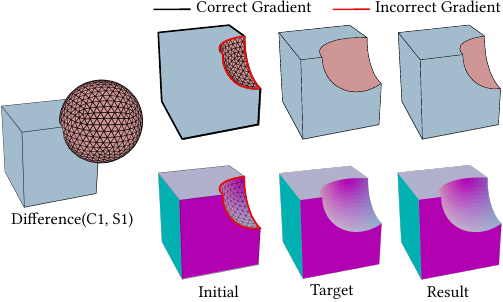}
    \caption{\textbf{Ablation study: impact of interior signal.} When primitives are rendered with solid colors, only silhouette and intersection edges provide signal to the optimization. If the intersection edges are not anti-aliased, the optimization fails (top).
    Rendering the primitives with interpolated vertex normals provides a richer visual signal within the spherical part, allowing the optimization to converge despite incorrect gradients at intersections (bottom). However, such interior signal only varies continuously for smooth primitives like the sphere, but is of no help for discontinuous primitives like the cube. Our full method, which applies anti-aliasing on intersection edges, succeeds with either solid colors or normals in that example.
    }
    \label{fig:abl_interior}
\end{figure}

\paragraph{Interior signals matter}
The success of the optimization is also influenced by the signal of the target image. As shown in Fig. \ref{fig:abl_interior}, interpolated vertex normals provide a richer internal signal than solid colors for curved surfaces, allowing the optimization to converge even when intersection edges are not properly anti-aliased. Nevertheless, vertex normals do not suffice to drive the optimization for piecewise flat surfaces, where large portions of the shape share the same orientation. 

\subsection{Comparison}
\label{subsec:comp}

\rev{
We have compared our method against two SOTA approaches for CSG model optimization.}

\begin{figure}[!htb]
    \centering
    \begin{overpic}[width=0.95\linewidth]{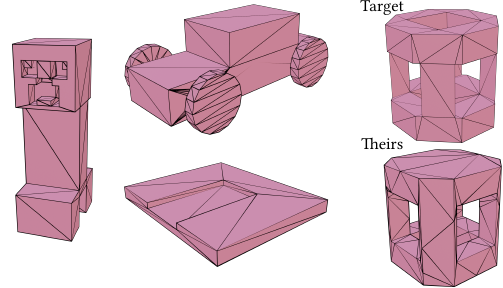} 
        \put(2,5) {\scriptsize Time: $1941.5s$ vs. $94.9s$}
        \put(2,1) {\scriptsize $S_{CD}$: $0.012$ vs. $0.015$}
        \put(47.5,33.5) {\scriptsize Time: $3131.1s$ vs. $74.3s$}
        \put(47.5,29.5) {\scriptsize $S_{CD}$: $0.013$ vs. $0.021$}
        \put(50,5) {\scriptsize Time: $1343.5s$ vs. $49.4s$}
        \put(50,1) {\scriptsize $S_{CD}$: $0.015$ vs. $0.019$}
    \end{overpic}
    \caption{\textbf{Comparison with the SDF-based differentiable CSG model from UCSG-Net} \cite{kania2020ucsg}. 
    The running time and $S_{CD}$ (theirs vs. ours) are shown for the first three examples, while their resulting shape for the last example is shown for comparison.
    \rev{Spefically, both methods were successful in the first three examples, achieving a similarly low Chamfer distance but at a very different computational cost. Since the difference between the resulting shapes is small, we only show the GT shapes. Our method succeeds for the last shape, but UCSG-Net fails to create a square hole in the center of the cylinder.}
    }
    \label{fig:timeCompare}
\end{figure}

\paragraph{SDF-based optimization - UCSGNet \cite{kania2020ucsg}.} \rev{Please refer to the supplemental material for the detailed configuration and analysis.
Briefly,} this experiment reveals that SDF-based optimization is \emph{three orders of magnitude} slower than our algorithm (see Fig.~\ref{fig:timeCompare} for typical cases), and fails to converge in several cases. Computational cost is due to the need to sample the SDF densely in 3D space, which exhibits cubic complexity compared to the quadratic complexity of our rendering-based approach.

\rev{
\paragraph{Derivative-free optimization - CMA-ES~\cite{hansen2003reducing}.} 
An alternative way of optimizing CSG shapes is utilizing derivative-free methods \cite{fischer2023plateau,deliot2024transforming,fischer2023zero,hansen2003reducing}, and we have compared with CMA-ES~\cite{hansen2003reducing}. 
For simple shapes with few parameters, CMA-ES can converge to the target state successfully, while it is hard to fit complex shapes with parameters of different and unknown scales. Please refer to the supplementary for a visual comparison and detailed analysis.
}

\subsection{Applications}
\label{subsec:application}

Our method can be applied to various inverse-rendering scenarios, we now introduce two representative examples.

\begin{figure}[!htb]
    \centering
    \includegraphics[width=0.98\linewidth]{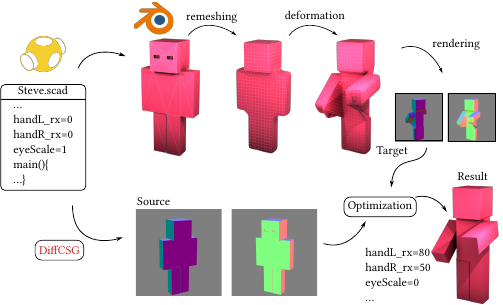}
    \caption{\textbf{3D editing.} The 3D mesh produced by a CSG model is exported to Blender for editing. In this example, we refined the mesh and deformed it to close the eyes and raise the arms. We then render normals of the edited mesh from two views to form the targets of our optimization, which successfully adjusts the size of the eyes and rotation of the arms to reproduce the edit.
    }
    \label{fig:3dEditng}
\end{figure}

\paragraph{Direct 3D editing}
Editing a CAD model via its parameters can be a tedious, trial-and-error process, which has motivated numerous approaches for direct editing of the resulting shape \cite{cascaval2022differentiable,Gonzalez2023,michel2021dag}. Fig. \ref{fig:3dEditng} illustrates how our approach can serve such a direct editing workflow. Given a CSG model with default parameters, the user can export the corresponding mesh, edit it in a 3D modeling software, and provide the edited mesh as the target for our optimization to find the CSG parameters that best reproduce that edit. In this example, the edit consists of refining and deforming the mesh in Blender, and two renderings of the result are used as the target for optimization, which takes $100$ seconds to converge.
Please refer to the supplemental video for the optimization process. 

\begin{figure}[!htb]
    \centering
    \includegraphics[width=0.99\linewidth]{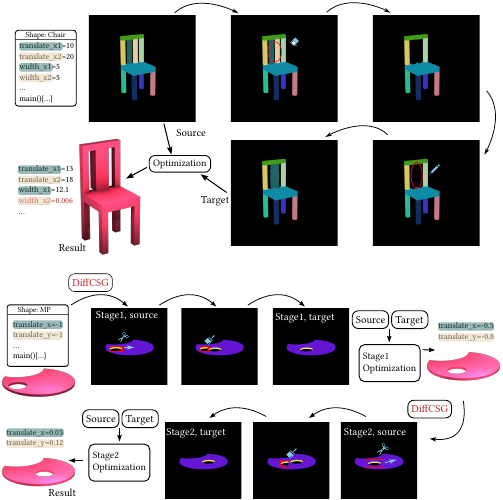}
    \caption{\textbf{Image-based editing.} Given a rendered color image of a 3D shape, direct editing on images can be mapped back to the 3D shape through our image-based differentiable optimization. Note that our optimization utilizes only a single image to perform these edits. 
    }
    \label{fig:imgediting}
\end{figure}

\paragraph{Image-based editing}
Our differentiable rendering approach also allows image-based editing of CSG models, where a rendering of the shape is loaded in an image editing interface to create a target for our optimization. 
Fig. \ref{fig:imgediting} demonstrates this workflow in two examples. In the first example, given a chair with four splats in the back, the user edits a rendering to erase two of the splats and draw a larger one instead.
The second example shows an iterative edit, where the user displaces a hole using image cutting and painting. The edit is performed in two steps to ensure that the holes overlap in the source and target images. Without any overlap, the optimization couldn't progress.

\subsection{Discussions}
\label{subsec:discussion}

\begin{figure}[!htb]
    \centering
    \begin{overpic}[width=0.98\linewidth]{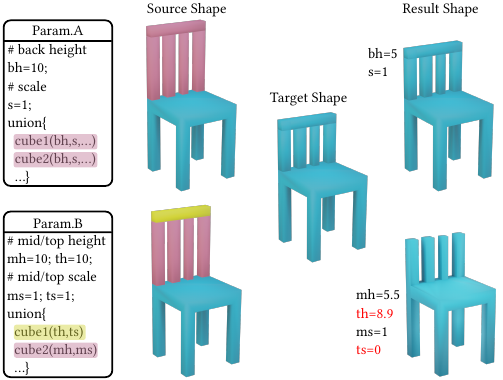}
    \end{overpic}
    \caption{\textbf{Parameterization variation.} Given the same target shape, co-parameterization of different components eases optimization as the top rail is constrained to remain attached to the splats (top), while independent parameterization yields a broken shape where the top rail disappears instead of moving to its target position (bottom).
    }
    \label{fig:hyper_params}
\end{figure}

\paragraph{Impact of CSG parameterization}
As discussed in Section \ref{subsec:dataset_evaluation}, CSG models are easier to optimize if their editable parameters are abstracted into a few hyper-parameters (see $Succ$ for H-Params and D-Params). 
Moreover, since the way to abstract hyper-parameters depends on expertise and design purposes, there exists different hyper-parameter abstraction schemes. Fig. \ref{fig:hyper_params} presents an interesting example. As can be seen, the splats of the chair back can be co-parameterized or parameterized independently of the top rail. Thus, given the same target shape, the chair with shared parameters can reach the target successfully, while the other one fails to retain the top rail.

\begin{figure}[!bt]
    \includegraphics[width=0.99\linewidth]{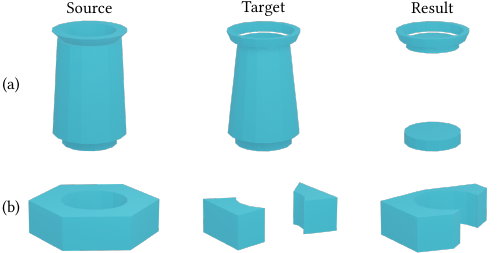}
    \caption{\textbf{Failure cases.} (a) In this example, the middle part is produced by subtracting a cylinder from another one. During the optimization, the subtracted cylinder gets a larger radius than the other one, making the entire part vanish. When a primitive vanishes, it does not produce any pixel, and as such does not produce any gradient to reappear.
    (b) The source nut shape is too far from the target. Despite optimization, the resulting shape only partially reproduces the target.}
    \label{fig:failure}
\end{figure}

\paragraph{Failure cases.}
Our differentiable optimization has two typical failure cases as shown in Fig. \ref{fig:failure}.
\begin{itemize}
    \item Vanishing primitives. It may happen that some of the primitive parameters (\eg the radius) are updated to be \emph{zero}, leading to the vanishing of specific primitives. However, since our method relies on visible pixels to obtain gradients, once a primitive disappears, there is no chance to bring that primitive back to the optimization (Fig. \ref{fig:failure}(a)). To address this problem, a potential solution is to insert a regularizer into the optimization to ensure that parameter values remain above zero.
    However, the ability to make primitives vanish can be a desired feature, such as when the initial shape contains redundant primitives. 
    \item Distant target shape. When the target shape is far from the source shape, the optimization may fail, as shown in Fig. \ref{fig:failure}(b). In such cases, a step-by-step iterative optimization is a possible solution to gradually reach the target shape.
\end{itemize}

\section{Conclusion and Future Work}

We have presented \name to differentiably render CSG models. We enable this by building upon the classical Goldfeather rasterization algorithm that displays the results of boolean operations on primitives \textit{without} explicitly computing the resultant mesh. Specifically, we demonstrate how to identify potential primitive intersections under specified boolean operations, and explicitly rasterize the intersection edges to propagate gradients back to the parameters of the input CSG program. Our method effectively marries meshing-free CSG rasterization and differentiable polygonal rasterization. 
The method is simple and easy to integrate into existing inverse rendering setups. 
We demonstrate the effectiveness and flexibility of our method on a newly introduced CSG optimization benchmark, as well as through image-based CAD editing applications.

\paragraph{Limitations and Future Work}

\begin{enumerate}
    \item {\em Structure Optimization for CSG Programs:} 
    Currently, our method cannot change the structure of the input CAD program. In the future, we would like to investigate program structure modifications (i.e., changing primitive types and modifying boolean operations). One possible direction is to combine our \name for continuous parameter changes with an RL-based search to optimize over discrete program changes. 
    \item {\em Effect of Initialization:} 
    While our main contribution is avoiding explicit meshing of CSG programs, and hence capturing gradients from possible intersection edges, \name can still end up in a local minima based on the starting state of the CSG program. Using a hypernetwork to provide an initialization for any CGS programs is a natural option to explore to address this issue. 
    \item {\em Regularizations:} 
    Our current implementation does not support any regularization as our contribution is to enable gradient flow, in the context of invisible edges arising from CSG operations, that is missed by existing differentiable mesh renderers. A natural next step would be to regularize the underlying programs by either using shape priors or restricting the shapes to stay in a latent space of (pretrained) CAD programs~\cite{jones2023ShapeCoder}. This can additionally allow us to perform both structure changes based on moving in the CAD program latent space as well as continuous parameter refinement using \name.
\end{enumerate}

\begin{acks}
\rev{
The authors would like to thank the reviewers for their valuable suggestions, Kuankuan Cheng and Shuyuan Zhang for helping prepare the benchmark and set up the comparison. NM was supported by Marie Skłodowska-Curie grant agreement No. 956585, gifts from Adobe, and UCL AI Centre; AB was supported by ANR-NSF NaturalCAD (ANR-23-CE94-0003); CJ was supported by a startup grant, a Bayes seed funding, and a GAIL seed funding from the University of Edinburgh, and gifts from Adobe.
}
\end{acks}

\bibliographystyle{ACM-Reference-Format}
\bibliography{diffCSGBib}


\end{document}